\documentclass[lettersize,journal]{IEEEtran}

\usepackage{xcolor,soul,framed} 
\usepackage{graphicx} 
\graphicspath{{../pdf/}{../jpeg/}} 
\DeclareGraphicsExtensions{.pdf,.jpeg,.png} 

\usepackage{array}
\usepackage{amsmath,amsfonts}
\usepackage[caption=false,font=normalsize,labelfont=sf,textfont=sf]{subfig}
\usepackage{textcomp}
\usepackage{stfloats}
\usepackage{times,epsfig}
\usepackage{mdwmath} 
\usepackage{mdwtab} 
\usepackage{eqparbox} 
\usepackage{url} 
\usepackage[font=footnotesize]{caption}
\usepackage{subcaption} 
\usepackage{xcolor}
\usepackage{soul}
\usepackage{dcolumn}
\usepackage{cite}
\usepackage{lipsum}

\begin{document}

    \title{Capacitive-Tuned SIW Evanescent-Mode Cavity for Resonant Microwave Plasma Jet}
  \author{Kazi~Sadman~Kabir,~\IEEEmembership{Graduate Student Member,~IEEE,}
      and~Abbas~Semnani,~\IEEEmembership{Senior Member,~IEEE}

  \thanks{The authors are with the Department of Electrical Engineering and Computer Science, The University of Toledo, Toledo, OH 43606 USA (e-mail: kazisadman.kabir@rockets.utoledo.edu; abbas.semnani@utoledo.edu). This paper is based upon work supported by the National Science Foundation under Grant No. 2102100.}
  }  

\markboth{IEEE TRANSACTIONS ON MICROWAVE THEORY AND TECHNIQUES,~VOL.~x, NO.~xx, 2024
}{Roberg \MakeLowercase{\textit{et al.}}: High-Efficiency Diode and Transistor Rectifiers}

\maketitle

\begin{abstract}
This paper introduces a novel atmospheric pressure and frequency-tunable microwave plasma jet utilizing evanescent-mode cavity resonator technology. The design uses a substrate-integrated waveguide approach, where two PCB substrates are assembled to form the resonant microwave plasma jet structure. This configuration provides excellent matching performance across the tuning range and can generate power-efficient plasma jets with a minimal input power of just a few watts. The prototype's resonant frequency can be tuned from 2.94 GHz, with no tuning capacitor, to 2.66 GHz when a capacitor placed in parallel with a circular slot etched on the top PCB board varies up to 1.6 pF. A novel circuit model design approach for this prototype is presented, and close agreement between the measured performance of the fabricated prototype and the simulation results validates this model. The frequency tunability achieved is a critical feature for fine-tuning the plasma jet when fabricated by low-cost PCB manufacturing and for applications in which manipulating the plasma chemistry by varying the ignition frequency is essential.
\end{abstract}

\begin{IEEEkeywords}
EVA Cavity Resonator, frequency tunable, plasma jet, substrate-integrated waveguide
\end{IEEEkeywords}

%
\IEEEpeerreviewmaketitle

\section{Introduction}

\IEEEPARstart{A}{tmospheric} plasma jets have emerged as a technology of significant interest in a broad spectrum of applications, including cancer therapies \cite{ct}, blood coagulation \cite{bc}, wound healing \cite{wh}, food sterilization \cite{fs}, water purification \cite{wp}, and electric propulsion \cite{pt}. These devices generate weakly ionized plasmas characterized by electron temperatures substantially exceeding the temperatures of surrounding heavy particles (T$_{i}$ $\approx$ T$_{g}$ $\approx$ 300 K < T$_{e}$ $\leq$ 10$^{5}$ K), thereby distinguishing them from high-temperature thermal plasma devices. Upon interaction with ambient air, cold plasma jets produce reactive oxygen and nitrogen species (RONS), including singlet oxygen ($^{1}$O$_{2}$), hydrogen peroxide (H$_{2}$O$_{2}$), superoxide anion (O$_2^{-}$), nitric oxide (NO), and ozone (O$_{3}$). These species play pivotal roles in various biomedical contexts. Through control of the plasma generation process, RONS profiles can be tailored to exhibit specific characteristics conducive to distinct applications.

The state-of-the-art cold plasma jet devices \cite{cpjd} face challenges related to efficiency, compactness, and cost-effectiveness. Additionally, existing devices typically operate at fixed frequencies, severely limiting their applicability. The integration of frequency tunability into cold plasma jet devices could introduce distinctive capabilities, including enabling precise manipulation of reactive oxygen and nitrogen species (RONS) over a broader range. This enhancement would significantly expand the device's utility across multiple domains, representing a significant advancement in cold plasma jet technology. This innovation has the potential for far-reaching impacts in agriculture, plasma medicine, and reconfigurable RF electronics.

In generating power-efficient plasma jets at atmospheric pressure, evanescent-mode cavity (EVA) resonators have demonstrated efficacy because of their capacity to concentrate and store electromagnetic energy. The high-quality factor (Q) exhibited by EVA resonators renders them suitable for facilitating low-power gas breakdown through structural modifications, thereby facilitating efficient plasma formation across various resonant frequencies. In 2022, the authors reported on developing a 2.45 GHz efficient resonant plasma jet device employing EVA cavity resonator technology \cite{Semnani} and elucidated its gas dynamics \cite{SemnaniGD}. This device was fabricated using CNC techniques and showed a plasma power efficiency of approximately 80\%. However, despite its promising attributes, the fabrication of such structures necessitates microprecision, rendering it a costly and intricate process. In addition, there was no mechanism for frequency tuning.

A printed circuit board (PCB) implementation of the structure, utilizing the substrate-integrated waveguide (SIW) technique, provides enhanced ease of manufacturability at a reduced fabrication cost, simultaneously harboring the potential for frequency tunability, which is essential for specific applications. Within a static plasma jet configuration, variations in gas flow rate, power input, and background composition influence plasma dynamics and the generated chemical species. The capacity to manipulate the operating frequency of a plasma jet represents an additional control parameter for modulating these processes, given that distinct chemical reactions manifest themselves at different temporal regimes corresponding to varying frequencies.

SIW-compatible tuning topologies often incorporate various mechanisms such as piezoelectric actuators \cite{pz1,pz2}, contactless tuners \cite{tuner1,tuner2,tuner3}, and varactor diodes \cite{vd1,vd2}. Piezoelectric actuators, while practical, exhibit significant power dissipation at higher frequencies, leading to phenomena such as hysteresis and drift over time. These effects can compromise the precision and stability of the device. Similarly, contactless tuners are suboptimal for power-efficient plasma sources due to radiation losses. In contrast, surface-mountable tuning schemes that utilize varactor diodes offer a promising alternative. Varactor diodes have successfully tuned various radio frequency (RF) structures implemented with PCB technology. For instance, Akash et al. \cite{vd1,vd2} demonstrated the tuning mechanism of a SIW EVA cavity using varactors. The compatibility of surface mountable varactor tuning with PCB technology ensures optimal performance, simplifies the assembly, and enhances cost-effectiveness. While varactors exhibit high quality factors at lower frequencies, rendering them suitable for filter applications, their performance deteriorates significantly at microwave frequencies. This limitation hinders their efficacy in generating power-efficient plasma jets within microwave-driven systems. 

This paper proposes an alternative approach that uses high-voltage DC capacitors to achieve frequency-tunable plasma jet operation. A proof-of-concept demonstration is presented, leveraging an SIW-EVA cavity resonator structure operating under atmospheric pressure conditions. Integrating a capacitor-based tuning topology demonstrates the potential frequency-tunability feature of this novel plasma jet. Section II delineates the design methodology, including simulation studies and circuit modeling of the proposed system. Subsequently, Section III elucidates the fabrication and assembly procedures, the comprehensive measurement results, and the corresponding discussions. Finally, Section IV encapsulates the findings in a conclusive summary, highlighting the significance and potential implications of the developed frequency-tunable SIW EVA cavity resonator-based plasma jet.

\section{Design Theory and Simulation}

\subsection {An SIW EVA Resonator Design}

\begin{figure}[!]
\centering
 \includegraphics[width=0.7\linewidth]{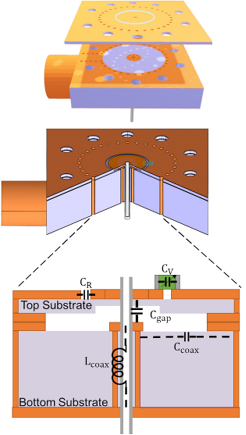}
   \caption{\footnotesize{The introduced frequency-tunable and SIW-based EVA cavity resonator using two different substrates to realize the cavity body (the bottom thick substrate) and its ceiling (the top thin substrate), which includes the critical gap area for plasma ignition. The main resonator parameters are highlighted on the structure.}}
   \label{fig:parameters}
\end{figure}

The proposed architecture comprises two distinct boards, as depicted in Fig. \ref{fig:parameters}, each fulfilling a specific function within the structure. The bottom board replicates the EVA cavity with critical elements, including the coaxial capacitance ($C_{coax}$) and the coaxial inductance ($L_{coax}$), dependent on parameters such as post radius, cavity radius, and substrate characteristics. The top board provides the cavity ceiling and primarily contributes to the formation of the gap capacitance ($C_{gap}$), spanning from the top of the post to the cavity ceiling, as shown in Fig. \ref{fig:parameters}. The $C_{gap}$ exhibits a pronounced sensitivity to environmental factors, which is practically addressed through a calibrated dielectric etching procedure, in which approximately 200 $\mu$m of substrate material is removed from the bottom side of the top board using a precision PCB milling machine. This etching ensures the symmetrical concentration of the $|E|$-field within the microgap area over the cavity post, thereby facilitating the generation of a power-efficient plasma jet when a gas flow is directed through a capillary tube through this concentrated $|E|$-field region. To sandwich the two substrates, screws are placed outside the cavity region, effectively bonding the boards and maintaining structural integrity. The design dimensions are outlined in Table \ref{parameters}.

{\renewcommand{\arraystretch}{1.2}
\setlength{\tabcolsep}{6pt}
\begin{table}
\centering
\caption{\label{tabble1}\uppercase{Parameters of the designed frequency tunable SIW-based EVA cavity jet}}
\begin{small}
	\begin{tabular} {c c c}
		\hline  \hline
		Parameter & Symbol & Dimension (mm) \\ 
		\hline
		cavity radius & b & 9.5\\ 
	  post radius & a & 1.6\\  
		slot radius & $r_{o}$ & 4.5\\
        slot width & $w_{o}$ & 0.5\\
        top substrate thickness & $h_{t}$ & 0.508\\
        bottom substrate thickness & $h_{b}$ & 6.35\\ \hline  \hline
	\end{tabular}
  \label{parameters}
\end{small}
\end{table}}

\subsection{Full-Wave Simulation Results}
\begin{figure}[!]
    \centering
    \includegraphics[width=0.92\linewidth]{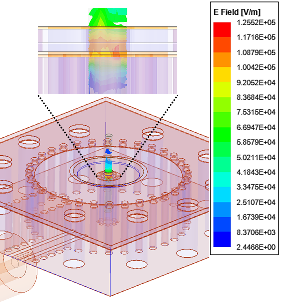}
    \caption{The simulated $|E|$-field of the SIW-based EVA cavity resonator over its critical gap area at the resonant frequency of 3.07 GHz and in the presence of a circular slot is very high, even with just 1 W of input power.}
    \label{fig:ef}
\end{figure}

The initial design exhibits a static resonant frequency of 3.07 GHz without including an external capacitor. For efficient plasma generation, a symmetrical and highly concentrated $|E|$-field is desired over the critical gap region. Previous studies of a static EVA plasma jet implemented on a copper structure have reported $|E|$-field values around 1.8$\times$$10^{5}$ V/m~\cite{SIWEVA}. However, in this tunable SIW design, the presence of the tuning slot and the associated dielectric losses because of SIW implementation of the design lead to a reduction in the $|E|$-field to approximately 1.25$\times$$10^{5}$ V/m, which is still sufficiently high for low-power gas breakdown and plasma jet formation. The concentrated electric field distribution observed in the simulations is shown in Fig. \ref{fig:ef} under 1 W of input power. This localized electric field facilitates gas ionization within the critical gap, initiating and sustaining plasma discharge.

\begin{figure}[!]
    \centering
    \includegraphics[width=.9\linewidth]{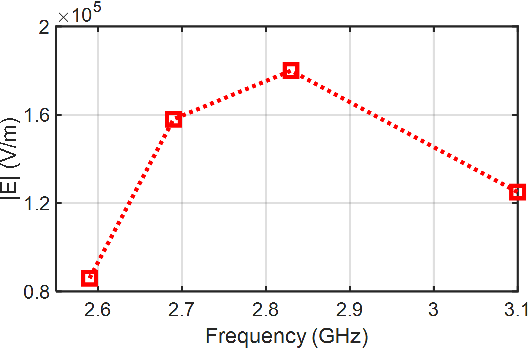}
    \caption{The simulated $|E|$-field of the SIW-based EVA cavity resonator at the tuned resonant frequencies}
    \label{fig:ef_tuned}
\end{figure}

The tunability of the resonant frequency of the SIW-based EVA cavity is achieved by introducing a circular slot on the top board of the cavity, as illustrated in Fig. \ref{fig:parameters}. The slot effectively creates a ring capacitance denoted as $C_{R}$. A $C_{V}$ capacitor is then placed parallel to $C_{R}$. Upon the capacitor's adjustment, the structure's overall equivalent capacitance is varied. This manipulation of the equivalent capacitance allows for corresponding adjustments to the resonant frequency. To model this tunability scheme in the simulations, $C_{V}$ is swept from 0 to 1.6 pF. The frequency
tuning performance obtained through the HFSS simulations is presented in Fig. \ref{fig:HFSSvADS}. As observed, further increases in capacitance resulted in a significant impedance mismatch, causing a substantial portion of the input power to be reflected. Hence, the $|E|$-field was studied at those tuned resonant frequencies as depicted in Fig. \ref{fig:ef_tuned} through simulation. The impedance mismatch reduces the $|E|$-field in the critical gap area, as low as 8.6$\times$$10^{4}$ V/m for $C_{V}$=1.6 pF in simulation. Hence, more power is required for plasma formation at that frequency. The breakdown and sustaining power conditions are discussed in detail in Section III.

\subsection{An Equivalent Circuit Model}
\begin{figure}[b]
    \centering
    \includegraphics[width=1\linewidth]{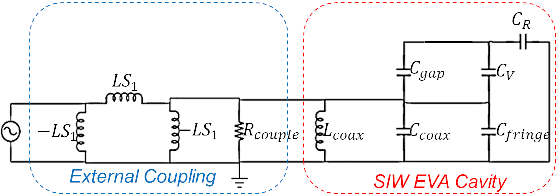}
    \caption{An equivalent circuit model of the proposed frequency tunable SIW-based EVA cavity resonator with the cavity and external coupling sections highlighted.}
    \label{fig:cm}
\end{figure}

A circuit model for the SIW-based EVA cavity is developed in two sections, representing the external coupling and the SIW EVA cavity itself, as illustrated in Fig. \ref{fig:cm}. The cavity excitation is performed by a 50-$\Omega$ coaxial cable connected to the EVA cavity through a coplanar-waveguide (CPW) line. This coupling scheme is captured in the circuit model using a $\pi$-inverter model, analogous to a coupling iris \cite{Chen2012}. The model elements $LS_1$ and $R_{couple}$ represent other external coupling parameters, which the following equations can describe \cite{excoupling}:
\begin{equation}
JS_{1}=M\left({\frac{\omega_{0}\, F_{bw}\, C_{eq}}{50}}\right)^{\frac{1}{2}},
\end{equation}
\begin{equation}
    \omega_{0}=\frac{1}{\left( {L_{coax}\, C_{eq}}\right)^{\frac{1}{2}}},
\end{equation}
\begin{equation}
LS_{1}=\frac{1}{\omega_{0}\, JS_{1}},
\end{equation}
\begin{equation}
    R_{p}=\frac{Q_{u}}{w_{0}\, C_{eq}}.
\end{equation}
Here, $C_{eq}$ represents the equivalent capacitance of the structure. $F_{bw}$ stands for the fractional bandwidth, and $Q_{u}$ is the unloaded quality factor of the resonator, which was extracted from the HFSS simulation to ensure optimal coupling between the 50 $\Omega$ line and the EVA resonator. The overall resonant frequency of the structure following empirical formulas is given by \cite{Semnani}
\begin{equation}
    f_{res}=\frac{1}{2\pi \sqrt{L_{coax}C_{eq}}}.
\end{equation}

The SIW EVA cavity section of the circuit model focuses on the resonator and its tuning mechanism. The elements $L_{coax}$, $C_{coax}$, $C_{gap}$, and $C_{fringe}$ represent the electrical characteristics derived from the structural dimensions and material of the cavity. The tuning mechanism, implemented by a circular slot etched onto the top substrate, introduces a ring capacitance, denoted as $C_{R}$. This capacitance can be estimated using \cite{tuner2}
\begin{equation}
\gamma=\frac{2\pi r_{0} \epsilon_{0}(1+\epsilon_{3})}{ln(1+\frac{w_{}}{r_{}})} \int_{0}^{\infty} [J_{0}(\zeta r)-J_{0}(\zeta(r+w_{}))]\frac{J_{1}(\zeta r)}{\zeta} \,d\zeta.
\end{equation}
Here, $\epsilon_{3}$ denote the dielectric constants of the medium below the slot, $w$ and $r$ represent the width and radius of the slot, respectively. The parameters $J_{0}$ and $J_{1}$ denote the Bessel functions of order 0$^{th}$ and 1$^{st}$, respectively. Following the introduction of the circular slot for tuning purposes, a high-voltage ceramic capacitor $C_{V}$ was surface mounted in parallel with the ring capacitance $C_{R}$ to achieve precise control over the structure's resonant frequency.

\begin{figure}[!]
    \centering
    \includegraphics[width=1\linewidth]{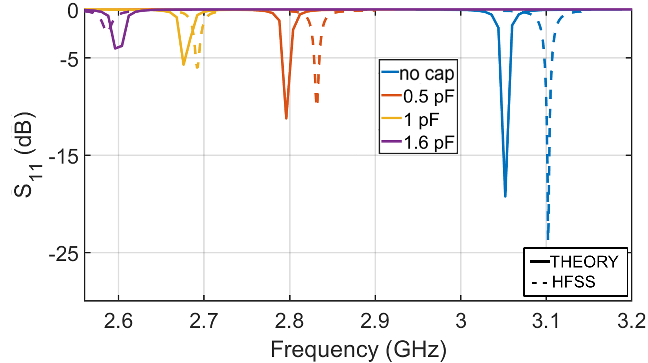}
    \caption{The equivalent circuit model versus HFSS full-wave simulated frequency tunability of the SIW-based EVA cavity resonator in pre-breakdown (plasma OFF) mode shows a good agreement.}
    \label{fig:HFSSvADS}
\end{figure}

The selection of the tuning capacitor values for practical implementation is guided by the circuit simulations performed in the Keysight Advanced Design System (ADS) simulator. Precisely, only capacitance values that result in reflection coefficients ($S_{11}$) below -10 dB are considered. Fig.~\ref{fig:HFSSvADS} compares the reflection coefficients predicted by the circuit model for various $C_{V}$ values and those obtained from full-wave simulations using ANSYS HFSS. To account for the additional etching at the bottom of the top substrate, the effective permittivity within the $C_{gap}$ region is incorporated into the ADS circuit model. With $\epsilon_{r}$ = 1.1 in the $C_{gap}$= {$\epsilon_{0}$$\epsilon_{r}$A/d}, the model resulted in a maximum frequency deviation of approximately 40 MHz for $C_{V}$ = 0 pF, compared to the HFSS simulation. Although the resonant frequencies predicted by the ADS model exhibited good agreement with the results obtained from the full-wave HFSS simulations, the observed frequency shifts can be attributed to high-frequency phenomena, such as fringe fields and slot radiation, which cannot be captured with a simple circuit model.

\section{Fabrication and Measurements}

Due to the unavailability of high-power and high-Q varactors with a wide tuning range, an alternative approach is explored to showcase the frequency tunability of the introduced microwave plasma jet device: The use of high-Q and high-voltage capacitors instead of varactors and the loaded quality factor ($Q_{l}$) was studied and compared with previous generations. These capacitors offer significantly a high Q-factor (at least 1,000) at microwave frequencies, mitigating the impedance mismatch issues encountered with varactors. In addition, they can handle voltages up to 1,000 V, making them ideal for high-power applications. In addition, the capacitor approach uses a single-slot design in which the introduced capacitance $C_{V}$ serves as the dominant capacitive element. This configuration translates to practical implementation with demonstrably good reflection coefficients and a significantly wider tunability range for the structure.

The unloaded Q-factor of the capacitor-less device is an important parameter that shows the ability of this device to efficiently store EM energy and is investigated through simulation and theoretical approaches. For a cylindrical evanescent mode cavity resonator, the unloaded quality factor can be given as \cite{qfactorEVA}
\begin{equation}
    \frac{1}{Q_{u}}=\frac{R_{S}}{2 \pi \mu f_{0}}\left( \frac{\frac{1}{a}+\frac{1}{b}}{ln\left(\frac{b}{a}\right)}+\frac{2}{h} \right).
\end{equation}
Based on the dimensions of the resonator introduced in Table \ref{parameters}, the calculated unloaded $Q$ is 12,694, which aligns well with the simulated $Q_u$ of 12,476 obtained using an HFSS eigenmode simulation. Such a high $Q_u$ guarantees significant enhancement over the resonator post, which is required for efficient gas breakdown and plasma formation. Hence, $Q_{l}$ was calculated from the measured $S_{11}$ and compared to previous generations of EVA resonators used to generate power-efficient plasma jets \cite{Semnani,SIWEVA}. \cite{Semnani} had a $Q_{l}$ of 612 with the metallic EVA resonator, which could sustain plasma as low as 500 mW. The transition from the metallic prototype to the SIW prototype reduced $Q_{l}$ to 214, which sustains plasma with a minimum input power of 2.8 W. This is understandable because the dielectric medium incorporates loss into the structure, unlike air. 

\begin{figure}[!]
\centering
 \includegraphics[width=1\linewidth]{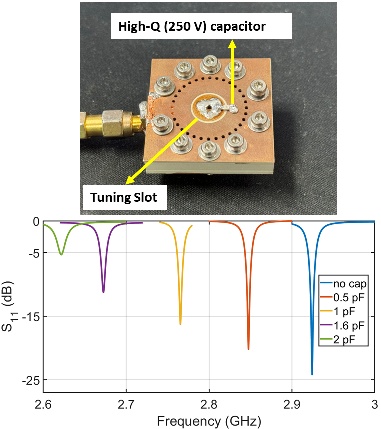}  
   \caption{Measured OFF-mode reflection coefficients of the frequency tunable SIW plasma jet for different capacitor values. Approximately a 300 MHz tuning range is observed.}
   \label{fig:Measured vs Simulated}
\end{figure}

For the frequency-tunable prototype, the reflection coefficient, $S_{11}$, was first evaluated for various capacitor values $C_{V}$ in the plasma OFF state, and subsequently, $Q_{l}$ was calculated from the measured $S_{11}$. Figure \ref{fig:Measured vs Simulated} presents the experimentally obtained $S_{11}$ responses for different $C_{V}$ values. The fabricated prototype exhibits a narrower tuning range compared to the simulation. This falls within the expected operational range but warrants further investigation into potential contributing factors. Several factors are believed to be responsible for the observed discrepancy between the simulated and experimental results.The calculated minimum $Q_{l}$ was 209 with capacitors at tuned resonant frequencies, showing a minimum $Q_{l}$ required to sustain power-efficient plasma jets in PCB prototypes with as low as 1 W of input power.

Following fabrication, an additional etching step is performed at the bottom of the top substrate using a PCB milling machine. This step aims to ensure a symmetric $|E|$-field concentration on the top of the post. However, slight inaccuracies could arise during this etching process due to inherent limitations in the precision of PCB milling machines. These deviations can influence the overall capacitance and, consequently, the resonant frequency. This observation suggests a potential deviation from the intended planar geometry during fabrication or etching. The PCB might have undergone a slight curvature in the central region, which deviates from the ideal straight-line profile. Furthermore, the prototype assembly involves manually drilling holes separate from the cavity vias to facilitate the sandwiching of the two boards with screws. Inconsistencies during this manual assembly process can introduce variations in the critical gap area. These dimensional changes directly affect the gap capacitance $C_{gap}$ and, consequently, the resonant frequency.

After the resonant behavior's characterization, the prototype's plasma ignition and jet profiles are investigated. A mass flow controller regulates the gas flow through the critical gap region. The length of the resulting plasma jet can be manipulated by adjusting this flow rate. Typically, plasma ignition occurs more efficiently at lower flow rates, with 1 slpm used. The flow rate is then gradually increased to observe the corresponding extension of the plasma jet. It should be noted that although Fig. \ref{fig:Measured vs Simulated} shows the tuning results with capacitors up to 2 pF, measurements are performed until the capacitance value of 1.6 pF has a reflection coefficient of better than -10 dB. In this case, the resonant frequency of the SIW EVA cavity prototype could be tuned to approximately 280 MHz when the capacitance of the surface mountable capacitor is varied up to 1.6 pF.

\begin{figure}[!]
\centering
   \includegraphics[width=0.93\linewidth]{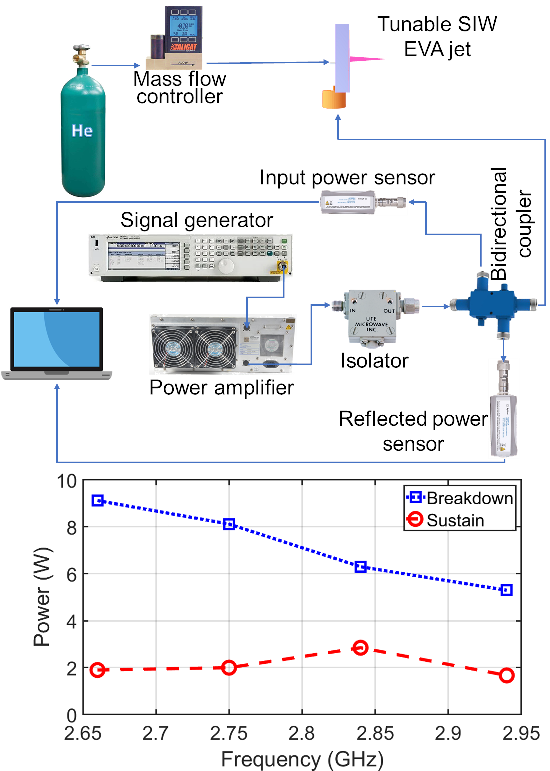}
   \caption{Measured plasma ignition and sustaining powers of the SIW EVA cavity resonator-based helium plasma jet at different resonant frequencies.}
   \label{fig:BdSt}
\end{figure}

In the next step, plasma formation and sustaining powers are evaluated at different resonant frequencies, with helium as a flowing gas, as depicted in Fig. \ref{fig:BdSt}. The breakdown power is the minimum power required for initial gas breakdown and plasma formation. Furthermore, the plasma sustaining power is the lowest power level at which stable plasma can be maintained. For this study, the helium flow rate is kept at 1 slpm at all frequencies as a low collision frequency among electrons is required for a low $|E|$ field gas breakdown. After plasma formation, the gas flow rate can be varied to achieve the desired plasma jet profile. With no capacitor mounted, the measured resonant frequency was 2.94 GHz, where plasma forms at 5.3 W. Usually, plasma sustains at lower power than the ignition power, which in this case was 1.67 W. This trend is consistent across all frequencies, as seen in Fig. \ref{fig:BdSt}.

\begin{figure*}[!]
    \centering
    \includegraphics [width=1\textwidth]{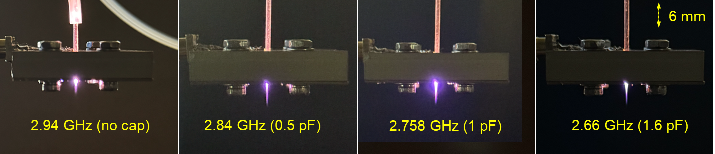}
    \caption{SIW EVA cavity resonator-based frequency tunable atmospheric plasma jet profiles at different frequencies, operated at 5 W input power and 7 slpm of helium flow rate.}
    \label{fig:Jet Profiles}
\end{figure*} 

After mounting capacitors and increasing their value, the structure's resonant frequency decreases and moves away from the optimized frequency of the no-capacitor case. Hence, the power required for gas breakdown and plasma formation increases because of higher reflection (worse matching), as depicted in Fig. \ref{fig:Measured vs Simulated}. However, the plasma sustaining power remains unchanged, averaging at 2 W over the entire frequency range, as seen in Fig. \ref{fig:BdSt}. This is among the lowest power consumption values for atmospheric pressure plasma jets.

Figure \ref{fig:Jet Profiles} presents the profiles of cold plasma jets generated at different resonant frequencies. For these measurements, the input power is kept constant at 5 W, and a mass flow controller (MFC) regulates helium flow rates at 7 slpm. Under these operating conditions, the maximum observed jet length is approximately 6 mm, almost consistent across all resonant frequencies tested. The maximum jet length is attained at a flow rate of around 4 slpm, providing the best helium laminar flow for the tube size employed. This is quantified using the dimensionless Reynolds number \cite{ReNu} defined as
\begin{equation}
    Re= \frac{\rho v d}{\eta},
\end{equation}
where $\rho$, \textit{d}, \textit{v}, and $\eta$ represent the density, diameter of the gas flow tube, speed, and viscosity, respectively. A Reynolds number of less than 2,000 indicates laminar flow when the gas travels smoothly and regularly. A higher Reynolds number indicates turbulent flow, which occurs when the gas flow is irregular and chaotic with a lot of mixing, resulting in shorter and less confined plasma jets.

The electron density ($n_{e}$) is a critical parameter of plasma jets and must be evaluated under different operating conditions. A passive spectrometry technique, known as optical emission spectroscopy (OES), is used to perform electron density measurements. This technique utilizes the spectral line profile and intensity at a particular wavelength. The two most common profiles used to analyze $n_{e}$ are the hydrogen-$\alpha$ line at $\sim$656 nm and the hydrogen-$\beta$ line at $\sim$486 nm, part of the spectral line emissions of hydrogen atoms. Gigosos and others widely used this method due to its convenience to estimate $n_{e}$ \cite{gigosos,ne2,ne3}. This method measures spectral full width at half-maximum (FWHM) at wavelengths directly related to the electron number densities. The spectral FWHM of hydrogen-$\alpha$ is chosen for this experiment because the line was stronger compared to hydrogen-$\beta$ and less sensitive to ion dynamics, making it a reliable value for the calculation of $n_{e}$ in gas discharges with a low degree of ionization \cite{ne}. The equation used to derive $n_{e}$ in cm$^{-3}$ from the FWHM of hydrogen-$\alpha$ line ($\lambda_{S}$) is given by \cite{gigosos}
\begin{equation}
    n_{e}=10^{17}\times\left(\Delta \lambda_{S}^{A}/1.098\right)^{1.47135}.
\end{equation}

\begin{figure}[!]
\centering
\includegraphics[width=0.85\linewidth]{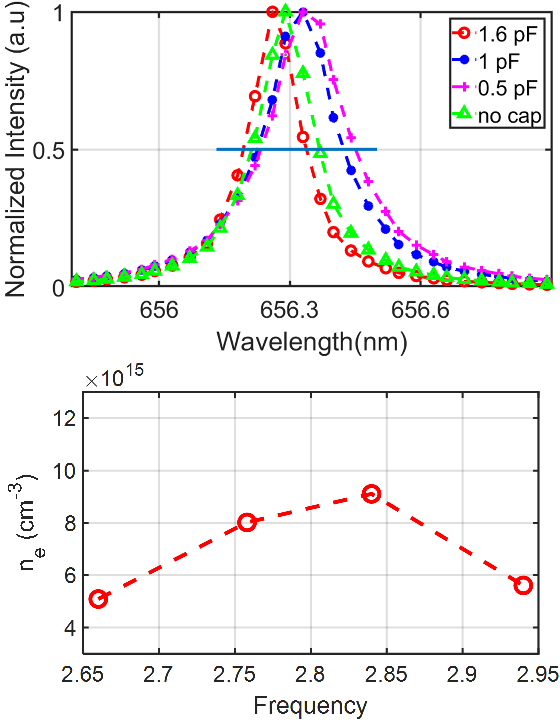}
   \caption{Top: FWHM Plot for 5 W input power and 1 slpm He flow rate using 0 - 1.6 pF capacitors to tune the resonant frequency. Bottom: Electron density of a frequency tunable SIW EVA cavity resonator-based helium plasma jet at different resonant frequencies.}
   \label{fig:Ne}
\end{figure}

Electron densities are measured by selecting 1, 5, and 10 W as input powers at a gas flow rate of 1 slpm. The gas flow rate is kept at 1 slpm as the plasma is confined at this flow rate, which provides a more accurate reading of the spectral profiles through a spectrometer with 0.05-nm spectral resolution. After measuring $n_{e}$ with input powers in the range of 1 to 10 W, it is observed that the effect of the input power on the electron density is negligible. Hence, in Fig. \ref{fig:Ne}, $n_{e}$ at different frequencies is displayed at only 5 W of input power. The $n_{e}$ shows the strongest values around 9.1$\times$10$^{15}$ cm$^{-3}$, where there is good impedance matching with a reflection coefficient of better than -15 dB. As the impedance mismatch increases with increasing $C_{V}$, the electron density drops to around 5$\times$10$^{15}$ cm$^{-3}$, which is still compatible with the $n_{e}$ range observed in static EVA-based plasma jets \cite{Semnani}. Higher plasma density can result in more interaction with the surrounding atmospheric air, triggering various chemical reactions that lead to the formation of a higher number of reactive oxygen and nitrogen species (RONS) production. Such RONS could be tailored for various applications, making it a suitable device for various plasma medicine applications.

{\renewcommand{\arraystretch}{1.2}
\setlength{\tabcolsep}{6pt}
\begin{table*}[!]
\centering
\caption{\label{tabble2}\uppercase{Performance comparison of Frequency tunable SIW Jet with state-of-the-art devices}}
    \begin{small}
	{\begin{tabular} {c c c c c c}
		\hline  \hline
		Device & Sustaining power  & Electron density  & Jet length & Frequency & Resonator \\ 
         & (W) & range $\left(cm^{-3}\right)$ & (mm) & tunability & type\\
		\hline
		\cite{Choi} & 0.5 & 10$^{14}$& 7 & No & Coaxial\\ 
	    \cite{Lee} & 2-10 & 10$^{15}$& 10 & No & Coaxial\\
        \cite{Porteanu} & 40 & 10$^{13}$ & N/A & No & Coaxial\\
		\cite{Semnani} & 0.5 & 10$^{15}$ & 1-6 & No & EVA cavity\\
        This work & 1.67 & 10$^{15}$ & 2-6 & Yes & SIW EVA cavity\\ \hline  \hline
	\end{tabular}}
  \label{comparison}
\end{small}
\end{table*}

The electron temperature is another plasma parameter that provides insight into the rates of chemical reactions within the plasma to generate various RONS. One widely used method of calculating the electron temperature is by observing the relative intensity of spectral lines emitted by plasma \cite{Te1}-\cite{Te5}. By observing the emission intensities of the helium line belonging to the same ionization state in an LTE plasma, the Boltzmann equation to estimate electron temperature $T_{e}$ is given by \cite{Te2} 
\begin{equation}
   T_{e}= - \frac{E_{k}-E_{i}}{k} \left[ln \frac{A_{k}g_{k}I_{i}\lambda_{i}}{A_{i}g_{i}I_{k}\lambda_{k}}\right]^{-1}.
\end{equation}
Here, $A_{i,j}$, $g_{i,j}$, and $E_{i,j}$ represent A-rates, degeneracies, and differences in the upper state energies of each chosen line, respectively, and these values were taken from the NIST website \cite{NIST}. $I_{i,j}$ is the line emission intensities of the two spectral lines observed through a spectrometer. The energy levels of the spectral lines 667.8 nm and 728.1 nm are resonant for the singlet, He states \cite{Te2}, and the resonant energy level has negligible quenching. Hence, these line intensities were used to calculate $T_{e}$ in the cold plasma generated through this device. 

The electron temperature of the device is calculated without any capacitor mounted at 2.94 GHz, as the device is optimized at that frequency. Fig. \ref{fig:Te} shows the electron temperature at different powers and helium flow rates within the laminar region of 1 to 2 slpm. The electron temperature is observed to increase with increased gas flow rate. With an increase in the gas flow rate, more neutral gas atoms collide with electrons and transfer energy to the electrons by heating them, which causes an increase in the electron temperature. 
\begin{figure}[!]
    \centering
    \includegraphics[width=0.9\linewidth]{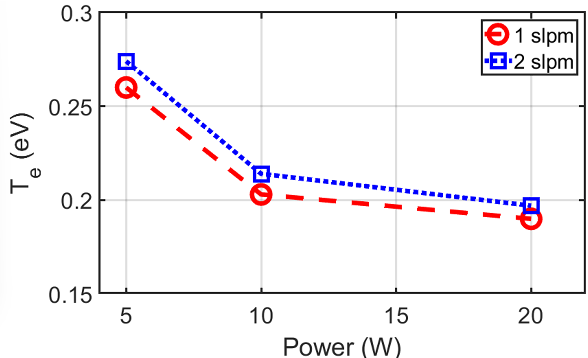} 
    \caption{Electron temperature of an EVA cavity resonator-based Helium plasma jet at different power and gas flow rates}
    \label{fig:Te}
\end{figure}

In contrast, the electron temperature decreases as the power supplied to the device increases. The device has a high power efficiency at lower input power. Hence, the lower the power provided to the device, the higher the portion of the input microwave energy is transferred to the plasma. This efficient input energy supply increases the kinetic energy of electrons, leading to a higher electron temperature. Understanding the electron temperature is crucial for gaining valuable information about the state of a plasma. This knowledge can then be used to determine the nature of chemical reactions and the generation of species that occur due to the plasma's interaction with its surroundings.

The tunable device presented in this work offers significant advancement over previous studies. Although other devices have demonstrated high electron densities, as Table \ref{tabble2} highlights, these solutions lack frequency adjustment. This is the first time such frequency tunability has been introduced in an atmospheric plasma jet device. It has provided a powerful tool for manipulating plasma for various applications such as materials processing, biomedical treatments, and environmental remediation. Furthermore, while many existing plasma jets focus solely on electron density and operational stability, the addition of frequency tunability in our design provides a performance advantage by enabling dynamic tuning of plasma jet physics and chemistry without sacrificing plasma stability or electron density, as demonstrated in our experimental results.

\section{Conclusion}
This work presents the first substrate-integrated waveguide-based and frequency-tunable resonant microwave plasma jet. The device consists of two PCB substrates that form a high-Q evanescent-mode cavity resonator. The resonant frequency is tuned using a capacitor mounted in parallel with a circular capacitive slot etched on the top of the upper board. The gas flow mechanism is facilitated by a capillary tube that passes through the center of the cavity. The device is compact and can achieve a high electron number density of approximately $\sim$$10^{15}$ cm$^{-3}$ with low input power in the range of a few watts. Achieving high electron number density with minimal input power enhances device's energy efficiency and versatility for various temperature-sensitive applications, including noninvasive biomedical ones. The frequency tunability also provides an additional mechanism to control the plasma's physical and chemical properties. 

\ifCLASSOPTIONcaptionsoff
  \newpage
\fi


\vfill

\end{document}